\documentclass[twocolumn,showpacs,preprintnumbers,prl,amsmath,amssymb,superscriptaddress]{revtex4}
\usepackage{graphicx}
\usepackage{endnotes}
\usepackage{bm}
\usepackage{epsfig}

\newcommand{\etal}{{\it et al.}} 
\newcommand{\BFAP}{BaFe$_2$(As$_{1-x}$P$_{x}$)$_2$}

\begin{document}


\title{Anisotropic Superconducting Properties of \\Optimally Doped BaFe$_2$(As$_{0.65}$P$_{0.35}$)$_2$ under Pressure}


\author{Swee~K.~Goh}
\email{skg27@cam.ac.uk}
\affiliation{Cavendish Laboratory, University of Cambridge, J.J. Thomson Avenue, Cambridge CB3 0HE, United Kingdom}

\author{Y.~Nakai}
\email{nakai@scphys.kyoto-u.ac.jp}
\author{K.~Ishida}
\affiliation{Department of Physics, Graduate School of Science, Kyoto University, Kyoto 606-8502, Japan}
\affiliation{TRIP, JST, Sanban-cho Building, 5, Sanban-cho, Chiyoda, Tokyo 102-0075, Japan}

\author{L.~E.~Klintberg}
\affiliation{Cavendish Laboratory, University of Cambridge, J.J. Thomson Avenue, Cambridge CB3 0HE, United Kingdom}

\author{Y.~Ihara}
\affiliation{Department of Physics, Graduate School of Science, Kyoto University, Kyoto 606-8502, Japan}

\author{S.~Kasahara}
\affiliation{Research Center for Low Temperature and Materials Sciences, Kyoto University, Kyoto 606-8502, Japan}

\author{T.~Shibauchi}
\author{Y.~Matsuda}
\affiliation{Department of Physics, Graduate School of Science, Kyoto University, Kyoto 606-8502, Japan}

\author{T.~Terashima}
\affiliation{Research Center for Low Temperature and Materials Sciences, Kyoto University, Kyoto 606-8502, Japan}

\date{\today}


\begin{abstract}
Magnetic measurements on optimally doped single crystals of \BFAP\ ($x\approx0.35$) with magnetic fields applied along different crystallographic axes were performed under pressure, enabling the pressure evolution of coherence lengths and the anisotropy factor to be followed. Despite a decrease in the superconducting critical temperature, our studies reveal that the superconducting properties become more anisotropic under pressure. With appropriate scaling, we directly compare these properties with the values obtained for \BFAP\ as a function of phosphorus content.\end{abstract}

\pacs{74.70.Xa, 62.50.-p, 74.25.Op} 

\maketitle

\section {Introduction}
The discovery of superconductivity in iron based compounds has triggered intense experimental and theoretical investigations \cite{Ishida_review}. One of the most heavily studied families is the so called 122 iron-arsenides (MFe$_2$As$_2$, where M=Ca, Sr, Ba and Eu), where the ground state can be tuned from a spin density wave (SDW) to superconductivity by charge doping \cite{Rotter08, Sasmal08} or pressure \cite{Alireza09, Terashima09}. For example, superconductivity in BaFe$_2$As$_2$ can be induced by hole-doping \cite{Rotter08}, electron-doping \cite{Ning09}, isoelectronic substitution \cite{Kasahara10}, and external physical pressure \cite{Alireza09}. 

The availability of high quality \BFAP\ single crystals has enabled detailed investigations of the charge and spin dynamics spanning the entire phase diagram. The SDW state is suppressed for $x\geq0.27$ and superconductivity is observed for $0.14\leq x\leq0.71$ where a maximum superconducting critical temperature, $T_c$, of $\sim$31~K is seen at $x\sim0.3$ \cite{Kasahara10, Chong10}. Quantum oscillation experiments revealed the existence of big Fermi surface sheets at the `overdoped' side ($x>0.33$) \cite{Shishido10}. In contrast, small Fermi pockets were detected in the parent compound of MFe$_2$As$_2$ \cite{Harrison09, Sebastian08}. In addition, an enhancement of the quasiparticle effective mass, $m^*$, was observed by quantum oscillations at the overdoped side on approaching the SDW state \cite{Shishido10}. On the other hand, nuclear magnetic resonance (NMR) measurements on the same part of the phase diagram reported a drastic enhancement of two-dimensional antiferromagnetic fluctuations on approaching the SDW state \cite{Nakai10}. Taking these experiments together as well as various anomalous transport properties \cite{Kasahara10}, it is evidenced that the \BFAP\ system features an antiferromagnetic quantum critical point at $x\sim0.33$. 

Apart from the interesting normal state properties mentioned above, the superconducting properties for $x=0.33$ also exhibit anomalous behavior. In particular, careful analysis of the temperature dependence of thermal conductivity \cite{Hashimoto10}, magnetic penetration depth \cite{Hashimoto10} and NMR spin-lattice relaxation rate \cite{Nakai10b} suggests the presence of line nodes in the superconducting gap.

The layered structure of the 122 systems naturally raises the question regarding the anisotropy of physical properties. The ability to apply magnetic field along both the ab-plane and the c-axis offers an excellent opportunity to study the anisotropy of superconducting properties under pressure. Here we report magnetic measurements for single crystals with $x\approx0.35$ under hydrostatic pressure. 

\section{Experimental}
\begin{figure}[!t]\centering
       \resizebox{8.5cm}{!}{
              \includegraphics{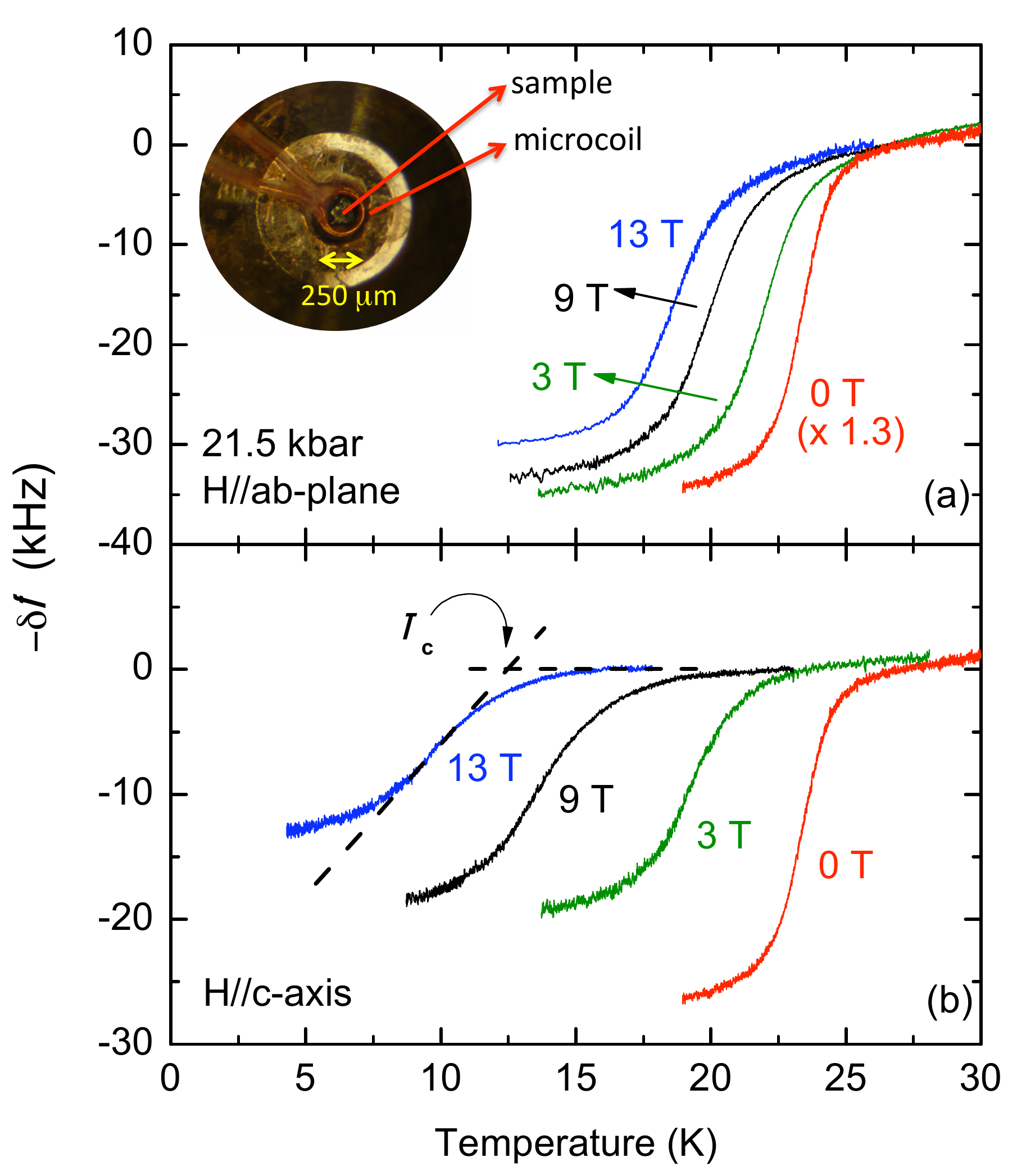}}               				
              \caption{\label{fig1} (Color online) Temperature dependence of the resonant frequency of the LCR circuit containing the sample in a microcoil (inset to (a)) at 21.5 kbar, with magnetic field applied along (a) the ab-plane and (b) the c-axis of the crystal. The traces were offset to the normal state value of $\sim$12~MHz.}
\end{figure}

High pressure experiments were performed using a Moissanite anvil cell on high-quality $x\approx0.35$ single crystals grown in Kyoto. The anvil cell is cylindrical with a diameter of 18~mm and a length of 40~mm, and can be rotated with respect to the magnetic field axis for anisotropy studies. Single crystals used for this study were grown from mixtures of FeAs, Fe, P(powders) and Ba (flakes) placed in an alumina crucible, sealed in an evacuated quartz tube and kept at $1150-1200\,^{\circ}\mathrm{C}$ for 12 hours, followed by a slow cooling to $900\,^{\circ}\mathrm{C}$ at the rate of $1.5\,^{\circ}\mathrm{C}$/hr \cite{Kasahara10b}. X-ray diffraction studies established the lattice constants to be $a=3.9266(12)$ and $c=12.8271(59)$, and the $z$ coordinate of pnictogen atoms in the unit cell is $z=0.35233(46)$, giving the phosphorus fraction of $x\approx0.35$. In addition, energy-dispersive X-ray spectroscopy established the variation in $x$ from sample to sample to be less than 2\%.  High sensitivity Faraday inductive measurements were performed by placing a 10-turn microcoil inside the gasket hole of the anvil cell \cite{Goh08, Alireza03, Haase09} (see the inset to Figure \ref{fig1}a). The superconducting transition was detected by two methods: (i) the shift in the resonant frequency of the LCR circuit consisting of a microcoil attached to an NMR probe (resonator mode), and (ii) a conventional mutual inductance method where a 140-turn modulation coil was also added inside the pressure cell. Glycerin was used as the pressure transmitting fluid and Ruby fluorescence spectroscopy was employed to determine the pressure achieved.

\section{Results and Discussion}
Figure \ref{fig1} presents the temperature dependence of the resonant frequency at different magnetic fields for $H \parallel$ ab-plane (Figure \ref{fig1}a) and $H \parallel$ c-axis (Figure \ref{fig1}b), for $x\approx0.35$ pressurized to 21.5~kbar. The transition to the superconducting state, manifested by an increase in the resonant frequency, is clearly observed. The superconducting critical temperature, $T_c$, at zero field and ambient pressure determined by this technique (not shown) is 30.5~K, which is in good agreement with the $T_c$ measured by the mutual inductance method (see the inset of Figure \ref{fig3}) and resistivity measurements reported elsewhere \cite{Kasahara10}. This value of $T_c$ confirms that $x\approx0.35$ is at the optimal doping of the \BFAP\ system. The superconducting transition, which is still very sharp at 21.5~kbar, demonstrates the high quality of the crystals and pressure transmitting fluid used in this study. 

\begin{figure}[!t]\centering
       \resizebox{9cm}{!}{
              \includegraphics{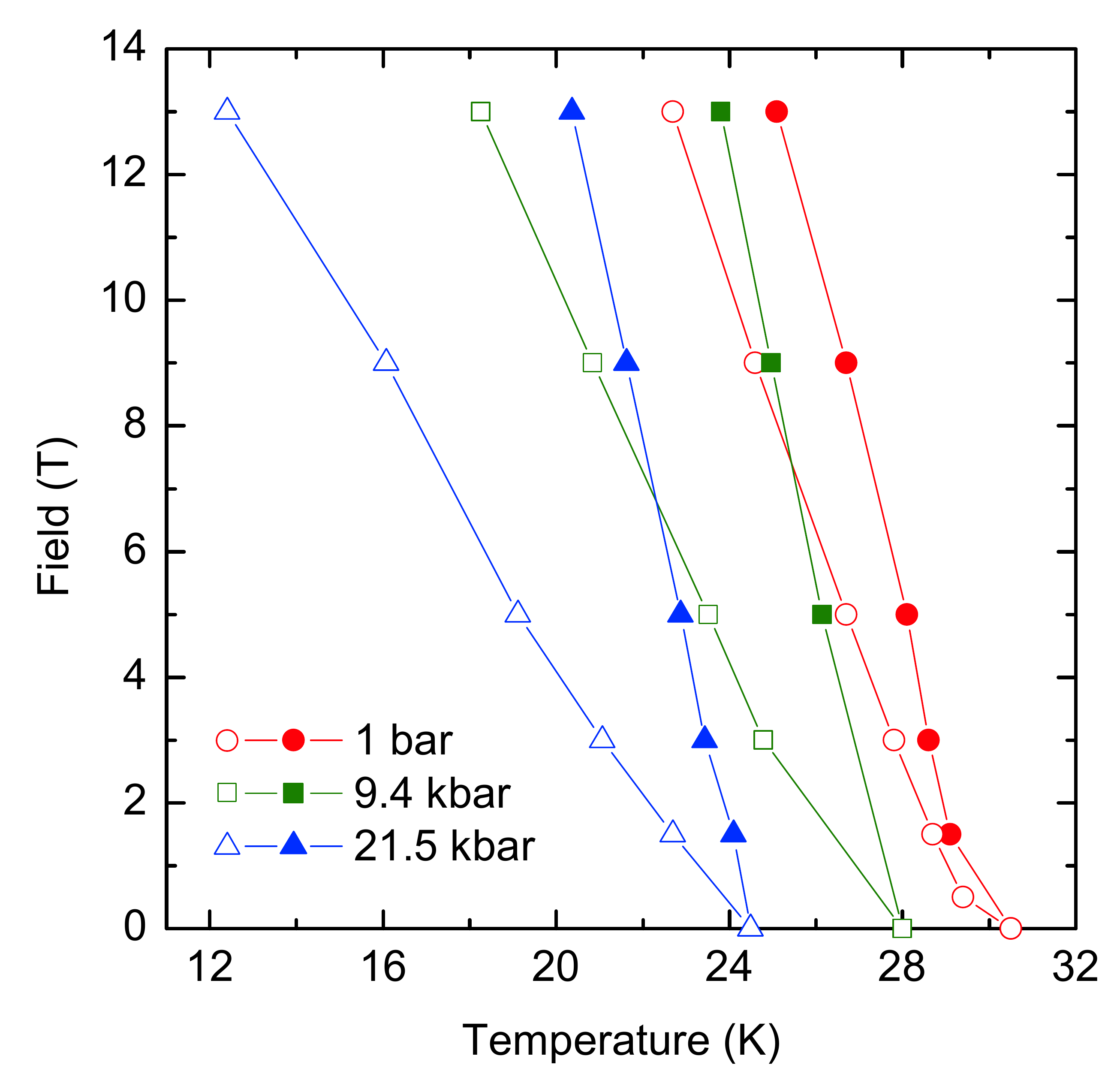}}               				
              \caption{\label{fig2} (Color online) Field dependence of $T_c$ for $H \parallel$ ab-plane (closed symbols) and $H \parallel$ c-axis (open symbols) at ambient pressure (circles), 9.4 kbar (squares) and 21.5 kbar (triangles)}
\end{figure}

\begin{figure}[!t]\centering
       \resizebox{8.5cm}{!}{
              \includegraphics{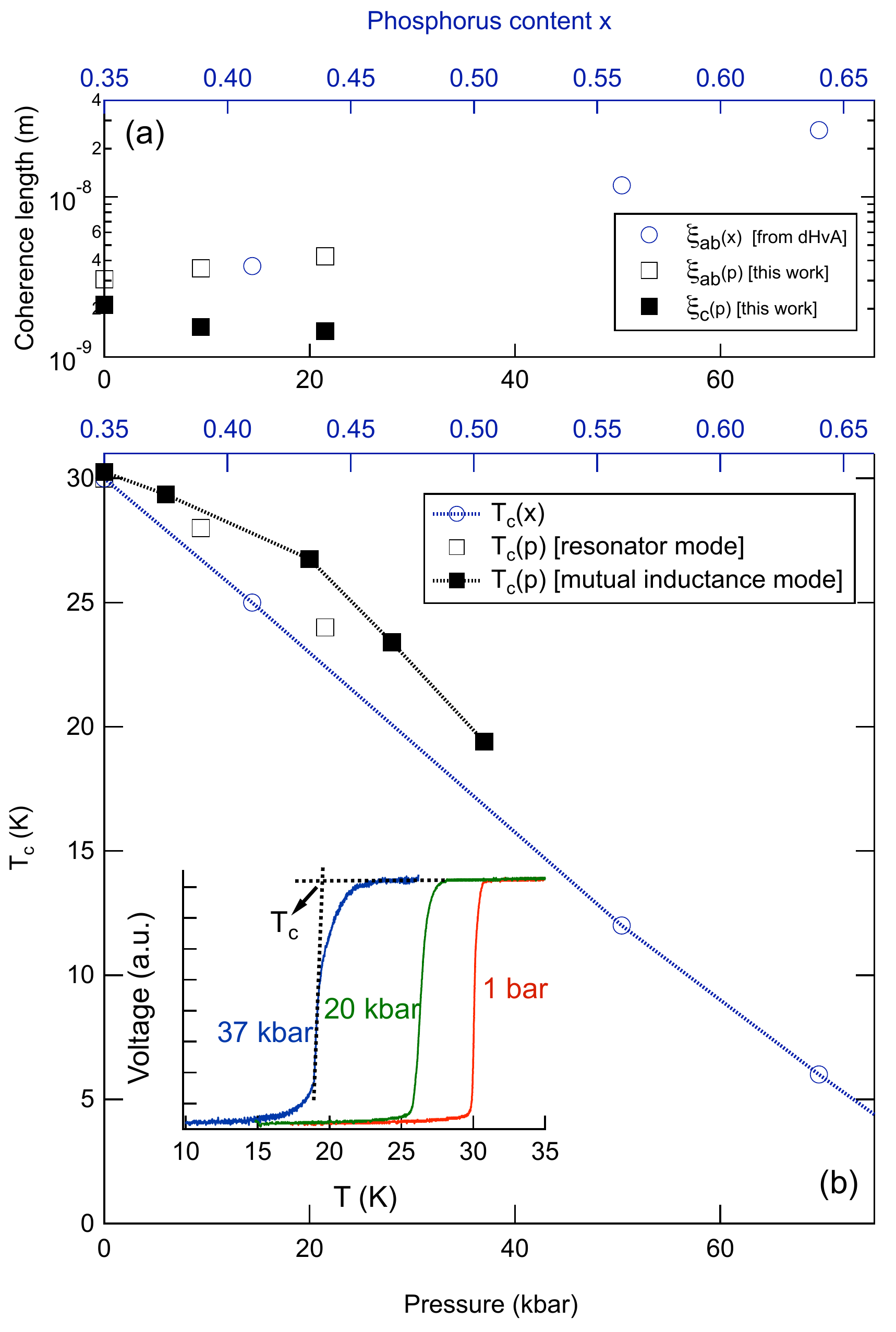}}               				
              \caption{\label{fig3} (Color online) (a) Coherence lengths extracted from the dHvA data\cite{Shishido10} (circles) and $H_{c2}$ (squares). (b) Pressure dependence of $T_c$ measured using the anvil cell in the mutual inductance mode (closed squares) and the resonator mode (open squares). $T_c(x > 0.35)$ (i.e. the overdoped part of the phase diagram) for \BFAP\ (upper axis) is also shown for comparison. The inset to (b) shows examples of traces obtained using the mutual inductance mode. Using the scaling relation where 1\% phosphorus corresponds to $\sim$2.4~kbar (see text), the phosphorus content is represented in both panels by the upper axis.}
\end{figure}

The superconducting transition with magnetic field applied shows a very distinct behavior for different field orientations. As evidenced in Figure \ref{fig1}, $T_c$ is depressed at a faster rate for $H \parallel$ c-axis. The $T_c$ values extracted at high fields are presented in Figure \ref{fig2} for ambient pressure, 9.4~kbar and 21.5~kbar. With increasing pressure, the rate of change of $T_c$ with field remains more or less constant for $H \parallel$ ab-plane whereas for $H \parallel$ c-axis, this rate increases strongly. Interestingly, high pressure resistivity studies on the optimally electron-doped Ba(Fe$_{1-x}$Co$_x$)$_2$As$_2$ \cite{Colombier10} and hole-doped Ba$_{1-x}$K$_x$Fe$_2$As$_2$ \cite{Torikachvili08} with $H \parallel c$ revealed only a small variation in the rate of change of $T_c$. Thus, the \BFAP\ system at optimal doping is rather special and we now extract anisotropic superconducting and normal state properties under pressure.

Apart from a slight curvature very near $T_c$ which has also been observed by other groups (see e.g. \cite{Butch10}), the variation of $T_c$ over our accessible field range is very linear. Equivalently, the initial slope of the $H_{c2}$--$T$ curves is approximately constant over the field range of 13~T. For ambient pressure data, $(dH_{c2}^{ab}/dT)|_{T_{c}}\sim-2.41$~T/K and $(dH_{c2}^{c}/dT)|_{T_{c}}\sim-1.67$~T/K for field along the ab-plane and c-axis, respectively.  According to the Werthamer-Helfand-Hohenberg (WHH) prescription \cite{WHH}, $H_{c2}(T\rightarrow0)=-0.69T_c(dH_{c2}/dT)|_{T_{c}}$, giving $H_{c2}^{ab}(0)\sim50.7$~T and $H_{c2}^{c}(0)\sim35.1$~T. We note that a similar WHH treatment by Chong \etal\ \cite{Chong10} based on resistivity data on BaFe$_2$(As$_{0.68}$P$_{0.32}$)$_2$ gives similar $H_{c2}^{c}(0)\sim36.4$~T but higher $H_{c2}^{ab}(0)\sim77.4$~T. In addition, Hashimoto \etal\ \cite{Hashimoto10} reported $H_{c2}^{c}(0)\sim51.5$~T based on specific heat and torque magnetometry measurements on BaFe$_2$(As$_{0.67}$P$_{0.33}$)$_2$. 

\begin{table}[!b]
\begin{center}
\begin{tabular}{c@{\hspace{0.2cm}} | @{\hspace{0.3cm}}c@{\hspace{0.3cm}} @{\hspace{0.3cm}}c@{\hspace{0.3cm}} @{\hspace{0.3cm}}c@{\hspace{0.3cm}}}   
\hline \hline
 & 1~bar   & 9.4~kbar   & 21.5~kbar  \\ 
\hline
$T_c (0~$T)\ (K) & 30.5 & 28.0 & 24.5  \\
$\xi_{ab}$\ ($\text{\AA}$) & 30.6&35.7 &42.6\\
$\xi_{c}$\ ($\text{\AA}$) & 21.2& 15.4&14.5\\
$\gamma$ & 1.44 & 2.32 & 2.93 \\
$m^*_{ab}(p)/m^*_{ab}(1$~bar) &1 &0.93 &0.90\\
$m^*_{c}(p)/m^*_{c}(1$~bar) &1 & 1.50&1.82\\
\hline \hline
\end{tabular}
\caption{\label{table1}Various physical properties extracted from $(dH_{c2}^{ab}/dT)|_{T_{c}}$ and $(dH_{c2}^{c}/dT)|_{T_{c}}$ for BaFe$_2$(As$_{0.65}$P$_{0.35}$)$_2$ at various pressures}
\end{center}
\end{table}

The anisotropy factor $\gamma$ is defined as $H_{c2}^{ab}/H_{c2}^c$, which for the same $T_c$ reduces to the ratio of the initial slopes $(dH_{c2}^{ab}/dT)|_{T_{c}}/(dH_{c2}^{c}/dT)|_{T_{c}}$. As tabulated in Table \ref{table1}, the pressure dependence of $\gamma$ is remarkable: relative to ambient pressure, $\gamma$ increases by more than a factor of 2 at 21.5~kbar despite a decreasing $T_c$. Since glycerin is hydrostatic at this pressure range, if we assume the anisotropy of superconducting properties is directly coupled to the anisotropy of the crystal lattice, it is tempting to conclude that the crystal lattice is more compressible along the $ab$ direction. However, neutron diffraction experiments performed on BaFe$_2$As$_2$ up to 60~kbar detected only minor changes in the lattice constants over the entire pressure range \cite{Kimber09}. Furthermore, the neutron result showed that the c-axis is slightly \textit{more} compressible than the $ab$ direction. For the heavy fermion superconductor CeCoIn$_5$, the anisotropy factor determined from the initial slopes decreases under pressure \cite{Knebel10}. This is accompanied by an initial increase in $T_c$, which peaks at 13~kbar. Furthermore, even more drastic behavior  is seen in CeCu$_2$Si$_2$: the anisotropy factor crosses $\gamma=1$ under pressure while $T_c$ increases strongly from 0.68~K at ambient pressure ($\gamma=1.24$) to 1.62~K at 24.2~kbar ($\gamma=0.87$) \cite{Sheikin00}. 

The Ginzburg-Landau coherence length along the two principal axes can be calculated from the set of equations \cite{Tinkhambook}

\begin{eqnarray}
	\xi_{ab}^2=\Phi_0/2\pi H_{c2}^c \\
	\xi_{ab}\xi_c=\Phi_0/2\pi H_{c2}^{ab}
\end{eqnarray}

\noindent where $\Phi_0=2.07\times10^{-15}$~Wb is the flux quantum. High pressure neutron diffraction \cite{Kimber09} found that the c-axis lattice constant decreases linearly by merely 2\% over 60~kbar. However, our analysis shows that $\xi_c$ already drops by 27\% at 9.4~kbar. This suggests that the coupling between the superconducting layers is significantly weakened upon applying a small amount of pressure. In contrast, $\xi_{ab}$ increases nearly linearly over the pressure range of our experiment.

BCS theory gives the (zero temperature) coherence length $\xi=\hbar v_F/\pi\Delta(0)=\hbar^2k_F/\pi m^*\Delta(0)$, where $v_F$ is the Fermi velocity, $k_F$ is the Fermi wavevector, $m^*$ is the effective mass and $\Delta(0) = a^*k_BT_c$ ($a^*=1.76$ in the BCS s-wave treatment) is the energy gap at zero temperature \cite{Tinkhambook}. This allows us to extract several normal state properties. Assuming that $k_F$ and $a^*$ are only weakly dependent on pressure, we estimate the \textit{relative} change of $m^*$ simply from the values of $T_c$ and $\xi$. In Table \ref{table1} we list the values of $m_i^*(p)/m_i^*(1$~bar) at pressure $p$ for two different principal axes $i=c$ and $i=ab$. $m_{c}^*$ increases with increasing pressure, implying a progressive reduction in the charge coupling along the c-axis. Interestingly, $m_{ab}^*$ decreases as pressure is applied, suggesting that pressure tunes the system away from a quantum critical point. This is analogous to the evolution of $m^*$ and $1/T_1T$ in \BFAP\ observed by the de Haas-van Alphen (dHvA) effect \cite{Shishido10} and NMR \cite{Nakai10b}, respectively, where a striking enhancement of both quantities has been observed on approaching the critical concentration $x\sim0.33$. We note that in CeRhIn$_5$, pressure $p>p_c=25$~kbar brings the system away from an antiferromagnetic quantum critical point \cite{Knebel10}. A similar analysis of the critical field data in CeRhIn$_5$ showed a reduction of the effective mass above $p_c=25$~kbar, the pressure at which $T_c$ is the maximum \cite{Knebel10}.

Bandstructure calculations for \BFAP\ indicate that the warping of the hole sheet becomes more pronounced with increasing $x$ \cite{Shishido10}. This suggests a more three dimensional Fermi surface with increasing $x$. If the Fermi surface also becomes more three dimensional under pressure, possibly due to a more compressible $c$-axis relative to the $ab$ direction (as observed by neutron diffraction on BaFe$_2$As$_2$ \cite{Kimber09}), the present observation of increasing anisotropy in superconducting properties is intriguing. One possible explanation is that the superconducting gap displays a significant $k_z$ dependence, becomes more anisotropic under pressure, and is smaller on the part of the Fermi surface where the Fermi velocity has a larger $k_z$ component \cite{Kogan02, Kogan03, Graser10}. Indeed, recent angle-resolved photoemission studies on the optimally hole-doped Ba$_{0.6}$K$_{0.4}$Fe$_2$As$_2$ found a significant $k_z$ dependence of the magnitude of the superconducting gap for a three dimensional hole sheet \cite{Zhang10, Xu10}.

Figure \ref{fig3}b shows the pressure dependence of $T_c$. Corroborating measurements were performed by the mutual inductance method. The initial variation of $T_c$ under pressure is approximately linear with the rate of $-0.18$~K/kbar. This rate is comparable to the initial suppression rate of about $-0.15$~K/kbar for the optimally hole-doped Ba$_{0.55}$K$_{0.45}$Fe$_2$As$_2$ \cite{Torikachvili08}, but slower than that for the optimally electron-doped Ba(Fe$_{0.926}$Co$_{0.074}$)$_{2}$As$_2$ (roughly $-0.25$~K/bar)\cite{Colombier10}. At 37~kbar, the transition is slightly broadened, possibly due to the inhomogeneity of the pressure transmitting fluid at this pressure. The fact that $T_c$ is depressed rather slowly and is still so high at 37~kbar is remarkable. This is to be contrasted to Alireza \etal's observation \cite{Alireza09} on BaFe$_2$As$_2$, where $T_c$ above $\sim$40~kbar is depressed rapidly with pressure, but remains rather constant above $\sim$50~kbar. 

The birth of superconductivity in the \BFAP\ system has been understood in terms of chemical pressure  exerted by phosphorus substitution. Comparing the initial slope of $T_{SDW}$ as a function of pressure \cite{Colombier09} against that for \BFAP\ at low phosphorus content $x$ \cite{Kasahara10}, we estimate that 1\% phosphorus content corresponds to $\sim$2.4~kbar. Using this linear scaling relation, the $T_c(x)$ can be added to Figure \ref{fig3}b for a direct comparison. For \BFAP\ at $x\sim0.33$, further increment of the phosphorus content depresses $T_c$. Since charge carriers are not introduced via phosphorus substitution, this reduction of $T_c$ has been attributed to the suppression of antiferromagnetic fluctuations \cite{Nakai10}, which provide glue for Cooper pairings. With physical pressure applied in our studies, which also does not introduce charge carriers, it is tempting to argue that a suppression of antiferromagnetic fluctuations is responsible for the reduction of $T_c$. The evolution of $m_{ab}^*$ discussed earlier hints at the possibility of this mechanism. NMR studies under pressure along this direction, which form the subject of future works, will be important to clarify this situation.

The dHvA experiment on \BFAP\ \cite{Shishido10} allows the calculation of $v_F(x)$ from the dHvA frequency via the Onsager relation \cite{Shoenbergbook}. Since for the dHvA experiment the magnetic field is applied along the c-axis, the in-plane coherence length, $\xi_{ab}$, can be extracted using the BCS theory discussed above. The value of the numerical constant $a^*$ for $\Delta(0)$ depends on the nodal structure of the gap as well as the dimensionality of the system \cite{Tinkhambook, Won94}. For our analysis, we take $a^*\approx2$. Figure \ref{fig3}a shows the $x$-dependence of $\xi_{ab}$ plotted on the same graph as $\xi_{ab}$ and $\xi_{c}$ under pressure, using the same pressure-phosphorus content scaling relation. Interestingly, taking into account only the Fermi velocity of the $\beta$ band, an electron band at the Brillouin zone corner, a smooth overall variation of $\xi_{ab}(p)$ and $\xi_{ab}(x)$ is observed. For a multiband system, the transition to the superconducting state is governed by the band with the shortest coherence length. Given that the band with the larger effective mass and the larger gap gives the shortest coherence length, this comparative study strongly suggests that the $\beta$ band might also have a larger gap, in addition to the large effective mass already observed \cite{Shishido10}. 

\section{Conclusions}
In summary, we have studied the effect of pressure on the superconducting properties of optimally doped \BFAP\ with $x\approx0.35$. We find that $T_c$ is suppressed rather slowly under pressure, and remains as high as 19.4~K at 37~kbar.  The coupling between superconducting layers becomes progressively weakened under pressure and the anisotropy in the superconducting properties \textit{increases} with applied pressure, which might be due to an increase in the anisotropy of the superconducting gap. The effective mass associated with the in-plane charge dynamics \textit{reduces} at higher pressures. With appropriate scaling, the pressure dependence of the superconducting properties agrees nicely with the $x$-dependence of the same properties for \BFAP. The $\beta$ band is likely to play a prominent role in determining the superconducting and thermodynamic properties of this system.
\begin{acknowledgments}

The authors acknowledge K. Shimizu, A. Miyake,  N. Naka, H. Hirabayashi for assistance with ruby fluorescence and M. Sutherland, G. G. Lonzarich, D. E. Khmelnitskii and P. L. Alireza for discussions. This work was supported by Grants-in-Aid for Scientific Research on Innovative Areas ``Heavy Electron'' (No. 20102006) from MEXT, for the GCOE Program``The Next Generation of Physics, Spun from Universality and Emergence'' from MEXT, and for Scientific Research from JSPS as well as Trinity College (Cambridge). S.K.G. acknowledges the Great Britain Sasakawa Foundation for travel grant and Kyoto University for hospitality. 

\end{acknowledgments}


\end{document}